\documentclass[technote,10pt]{IEEEtran}
\usepackage{amsmath,amsfonts,amssymb}
\usepackage{algorithmic}
\usepackage{algorithm}
\usepackage{array,booktabs,multirow}
\usepackage[caption=false,font=normalsize,labelfont=sf,textfont=sf]{subfig}
\usepackage{textcomp}
\usepackage{stfloats}
\usepackage{url}
\usepackage{tabularx}
\usepackage{array}

\newcolumntype{Y}{>{\centering\arraybackslash}X}
\usepackage{verbatim}
\usepackage{graphicx}
\usepackage{xcolor}
\definecolor{bestred}{RGB}{190,0,0}
\definecolor{secondblue}{RGB}{0,76,153}
\newcommand{\bestresult}[1]{\textcolor{bestred}{#1}}
\newcommand{\secondresult}[1]{\textcolor{secondblue}{#1}}
\usepackage[normalem]{ulem}
\usepackage{hyperref}
\usepackage{cite}
\usepackage{setspace}

\begin{document}

\bstctlcite{IEEEexample:BSTcontrol}

\title{%
TokenComSR: Task-Sensitivity-Guided Token Communication for Wireless Image Super-Resolution}

\author{Ye Wang, Li Qiao, Zhen Gao, and Hua Wang
\thanks{Ye Wang and Hua Wang are with the School of Information and Electronics, Beijing Institute of Technology (BIT), Beijing 100081, China (e-mail: bit\_wy@bit.edu.cn; wanghua@bit.edu.cn).

Li Qiao is with the Department of Electrical and Computer Engineering, The University of Hong Kong, Pokfulam Road, Hong Kong.

Zhen Gao is with Beijing Institute of Technology (BIT), Zhuhai 519088, China, also with the State Key Laboratory of CNS/ATM, Beijing 100081, China, and also with the MIIT Key Laboratory of Complex-Field Intelligent Sensing, Beijing 100081, China.

Corresponding authors: Li Qiao and Zhen Gao (e-mail: qiaoli@hku.hk; gaozhen16@bit.edu.cn).}}

\maketitle

\begin{abstract}

For resource-constrained wireless edge devices over bandwidth-limited fading channels, wireless image transmission using traditional separate  coding suffers from the cliff-effect collapse. Prevailing deep joint source-channel coding (JSCC) based on convolutional neural networks can mitigate this issue but usually fail to preserve patch-level structures, thereby preventing adaptive per-token power allocation and limiting token-domain compensation for super-resolution (SR). To address these challenges, we propose a token communication framework with SR (TokenComSR). Specifically, we conceive a task-sensitive power allocation (TSPA) module and a signal-to-noise ratio (SNR)-conditioned token refinement module (TRM). TSPA distills training estimates of task sensitivity into inference token power weights, while TRM estimates an SNR-conditioned residual to correct channel-induced distortion in the token domain before decoding. Building on TSPA and TRM, the proposed TokenComSR pairs a Swin Transformer-based token transceiver with a receiver-side SR module for resource-constrained wireless image transmission. Simulation results confirm the effectiveness of the proposed TSPA and TRM, demonstrating improvements over separate coding and JSCC-SR baselines in both reconstruction fidelity and perceptual quality.

\end{abstract}


\section{Introduction}
Resource-constrained edge devices, such as fixed roadside cameras and sensing nodes, lack onboard compute for local high-resolution (HR) image processing and offload reconstruction to resource-rich servers by transmitting compact low-resolution (LR) observations \cite{Ozer2023Offloading}. Such cloud--edge collaboration also supports integrated perception, communication, and computation in 6G networks \cite{li2025large}. Over fading, bandwidth-limited uplinks, however, channel distortion corrupts these representations before reaching the server and propagates through the reconstruction pipeline, fundamentally limiting HR output quality.

Conventional separate coding suffers from the \textit{cliff effect}, in which quality collapses abruptly once the channel signal-to-noise ratio (SNR) drops below a design threshold, making it ill-suited to varying wireless environments. Semantic communication (SC) has emerged as an enabling paradigm for edge intelligence \cite{Yang2022SemComEdge}, while deep joint source--channel coding (JSCC) mitigates this cliff effect through continuous source-to-channel mappings \cite{deepjscc2019}. Recent SC studies have further explored channel-adaptive semantic generation and interference-aware transmission \cite{yan2024adaptive}, as well as receiver-side generative reconstruction for perceptual quality enhancement \cite{zhang2025semantic}. Under the compute asymmetry of the edge-to-server setup, pairing JSCC with a receiver-side super-resolution (SR) network is natural: the resource-constrained sensing node transmits compact LR representations while the server reconstructs the HR output \cite{ref_sr_survey}. However, different image regions contribute unequally to downstream HR reconstruction, leading to heterogeneous protection requirements for their corresponding latent tokens. Existing JSCC-SR works \cite{jscc_sr} and adaptive JSCC methods such as ADJSCC \cite{adjscc} rely on convolutional neural network (CNN) encoders whose spatially entangled feature maps lack explicit support for token-level unequal protection or targeted pre-decoding refinement. This mismatch motivates allocating a fixed power budget across tokens according to their influence on HR reconstruction.

Recent advances in SC have motivated token communication (TokenCom), which organizes semantic information into addressable token representations aligned with the tokenized structures of generative and multimodal foundation models, thereby providing a structured interface for semantic-aware transmission and reconstruction \cite{qiao2025tokcom,qiao2024latency,yin2026generative,wei2026token,men2026video,qiao2025token}. For image-oriented TokenCom, the one-to-one correspondence between token and patch enables per-token power allocation and targeted pre-decoding correction. The same patch-structured attention also underlies state-of-the-art SR networks such as Swin2SR \cite{swin2sr}, indicating a natural architectural synergy between token transmission and receiver-side SR. Although SwinJSCC \cite{swinjscc} demonstrates Swin-Transformer (Swin-T)-based JSCC and prior TokenCom works address multi-perspective token-domain transmission, the explicit connection between downstream HR-task sensitivity, physical token power, and pre-decoding token repair has received limited attention, motivating the present work.

This paper proposes \textit{TokenComSR}, a TokenCom framework that integrates JSCC with learned SR under the resource-constrained edge-to-server paradigm. Exploiting the token-patch correspondence of the encoder, TokenComSR introduces task-sensitive power allocation (TSPA) at the transmitter and an SNR-conditioned token refinement module (TRM) before image decoding at the receiver, together with a progressive training strategy, enabling E2E optimization of both communication robustness and HR reconstruction quality. 

The main contributions of this paper are summarized as follows:
\begin{itemize}
    \item We develop TSPA, which distills downstream task sensitivity into token-level unequal power allocation under fixed channel usage and average-power constraints.
    \item We design TRM as an explicit pre-decoding token-domain repair mechanism that suppresses residual channel noise and restores cross-token relationships before nonlinear decoding and SR.
    \item We develop a progressive four-phase optimization strategy that coordinates the cascaded pipeline through staged activation and dedicated loss combinations, enabling stable E2E convergence.
\end{itemize}

\textit{Notation}: Lowercase, bold lowercase, and bold uppercase letters denote scalars, vectors, and matrices, respectively. $(\cdot)^*$, $(\cdot)^{\mathsf T}$, and $\operatorname{sg}(\cdot)$ denote complex conjugation, transpose, and stop-gradient, respectively.

\section{System Model}
\label{sec:system_model}

\begin{figure*}[!t]
    \centering
    \includegraphics[width=0.92\textwidth]{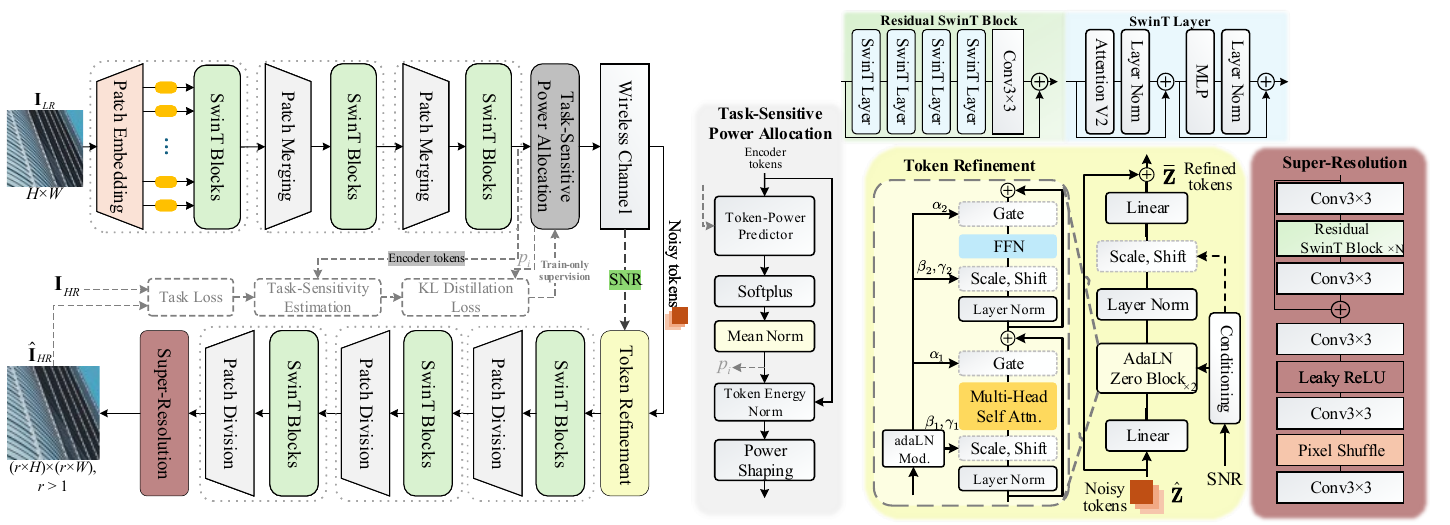}
    \caption{Overall architecture of the proposed TokenComSR. The LR image is encoded into semantic tokens. TSPA performs token-level power allocation and average-power normalization before transmission. The gray path denotes a training-only teacher branch that guides TSPA to score tokens according to their downstream HR-reconstruction sensitivity. At the remote server, TRM performs SNR-conditioned token refinement, after which the decoder and the SR module reconstruct HR output with upscaling factor $r$.}
    \label{fig:system_model}
\end{figure*}
Fig.~\ref{fig:system_model} shows the E2E TokenComSR pipeline. Let $\mathbf I_{\mathrm{LR}}\in\mathbb{R}^{H\times W\times3}$ denote the LR observation and $\mathbf I_{\mathrm{HR}}\in\mathbb{R}^{rH\times rW\times3}$ its HR target, where $r$ is the SR factor. The transmitter encoder maps $\mathbf I_{\mathrm{LR}}$ into $N$ latent tokens,
\begin{equation}
\mathbf Z=\mathcal E_{\phi}(\mathbf I_{\mathrm{LR}};\rho)
=[\mathbf z_1,\ldots,\mathbf z_N]^{\mathsf T}\in\mathbb R^{N\times C},
\label{eq:tokenization}
\end{equation}
where $C$ is the real dimension of each token. TSPA assigns the mean-normalized power coefficient
\begin{equation}
p_i=N\frac{q_i}{\sum_{j=1}^{N}q_j},
\qquad
p_i>0,\quad
\frac{1}{N}\sum_{i=1}^{N}p_i=1,
\label{eq:tspa_power}
\end{equation}
where $q_i>0$ is the predicted token score. The predictor and its task-sensitivity supervision are detailed in Section~\ref{sec:framework}.

Before transmission, each token is energy-normalized then scaled by $\sqrt{p_i}$, and mapped into complex channel symbols as
\begin{equation}
\tilde{\mathbf z}_i=\frac{\mathbf z_i}{\sqrt{C^{-1}\|\mathbf z_i\|_2^2+\epsilon}},
\qquad
\mathbf X=\xi\,\mathcal M\left(\left\{\sqrt{p_i}\,\tilde{\mathbf z}_i\right\}_{i=1}^{N}\right),
\label{eq:power_allocation}
\end{equation}
where $\mathcal M(\cdot)$ packs every two real-valued entries into one complex symbol, yielding $n_{\mathrm c}=NC/2$ channel uses per image. The scaling factor $\xi$ normalizes each image codeword to unit average power,
\begin{equation}
\frac{1}{n_{\mathrm c}}\left\|\mathbf X\right\|_2^2
=1.
\label{eq:average_power}
\end{equation}

Because the communication objective is to reconstruct the HR target, we define CBR relative to the number of RGB samples in the target image,
\begin{equation}
\mathrm{CBR}
=\frac{n_{\mathrm c}}{3r^2HW}
=\frac{NC}{6r^2HW}.
\label{eq:hr_cbr}
\end{equation}
This definition quantifies channel usage relative to the complete HR image. Unless otherwise stated, CBR denotes $\mathrm{CBR}_{\mathrm{HR}}$ throughout this paper.

For the considered fixed edge-sensing uplink, the wireless channel is modeled as quasi-static Rician block fading,
\begin{equation}
\mathbf Y=h\mathbf X+\mathbf N,
\label{eq:block_channel}
\end{equation}
where
\begin{equation}
h=
\sqrt{\frac{K}{K+1}}
+\sqrt{\frac{1}{2(K+1)}}
(g_{\mathrm R}+{\rm j}g_{\mathrm I}),
\label{eq:rician}
\end{equation}
$K\geq0$ is the linear ratio of the line-of-sight power to the scattered power, $g_{\mathrm R},g_{\mathrm I}\sim\mathcal N(0,1)$, and the entries of $\mathbf N$ are independently distributed as $\mathcal{CN}(0,N_0)$, with$N_0=10^{-\rho/10}$.

One realization of $h$ is shared by all symbols of an image and independently resampled across images; $K=0$ gives the Rayleigh limit. Thus, TSPA provides unequal token protection under a common channel attenuation rather than exploiting token-selective fading. During training, we sample several $K$ values to cover different LoS-to-NLoS power ratios, with the same values used for evaluation in Section~\ref{sec:experimental_settings}. 

After channel transmission and real--complex unpacking, the received tokens are denoted by $\hat{\mathbf Z}$. The receiver performs token refinement followed by cascaded image decoding and SR reconstruction as
\begin{align}
\bar{\mathbf Z}&=\mathcal T_{\omega}(\hat{\mathbf Z};\rho)
=\hat{\mathbf Z}+\Delta_{\omega}(\hat{\mathbf Z},\rho),\label{eq:trm_map}\\
\hat{\mathbf I}_{\mathrm{HR}}&=
\mathcal G_{\theta,\psi}(\bar{\mathbf Z};\rho),
\qquad
\mathcal G_{\theta,\psi}\triangleq
\mathcal R_{\psi}\circ\mathcal D_{\theta}.
\label{eq:hr_reconstruct}
\end{align}

\begin{figure}[t]
    \centering
    \includegraphics[width=0.93\linewidth]{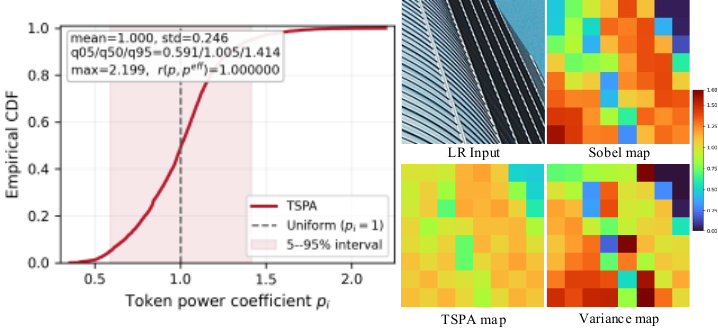}
    \caption{Empirical distribution and spatial maps of token power allocation for the same LR input.}
    \label{fig:tspa}
\end{figure}
\section{Proposed TokenComSR Framework}
\label{sec:framework}

\subsection{Task-Sensitive Power Allocation (TSPA)}

The TSPA predictor is a lightweight token-wise multilayer perceptron consisting of layer normalization, a bottleneck linear layer, GELU activation, and a scalar output layer. For each encoder token, it produces a positive score
\begin{equation}
g_i=f_{\eta}(\mathbf z_i),\qquad
q_i=\operatorname{softplus}(g_i)+\epsilon .
\label{eq:tspa_predictor}
\end{equation}
The scores are converted into the mean-one power coefficients $p_i$ through~(\ref{eq:tspa_power}). During training, a task-sensitivity teacher provides the target distribution for the predictor. Let $\mathcal L_{\mathrm{HR}}$ denote the HR-domain $L_1$ loss defined in Section~\ref{sec:training}. The first-order sensitivity of token $\mathbf z_i$ is
\begin{equation}
s_i=\left\|
\mathbf z_i\odot
\frac{\partial\mathcal L_{\mathrm{HR}}}{\partial\mathbf z_i}
\right\|_1 .
\label{eq:sensitivity}
\end{equation}
Under a fixed total-power budget, the first-order sensitivity surrogate gives
\begin{equation}
\min_{\substack{p_i>0\\\sum_{i=1}^{N}p_i=N}}
\sum_{i=1}^{N}\frac{s_i+\epsilon}{p_i}
\quad\Longrightarrow\quad
p_i\propto\sqrt{s_i+\epsilon}.
\label{eq:sensitivity_surrogate}
\end{equation}
This relation defines the teacher weights under the first-order surrogate. Accordingly, the teacher and predicted allocation distributions are
\begin{equation}
\pi_i=\frac{\sqrt{s_i+\epsilon}}
{\sum_{j=1}^{N}\sqrt{s_j+\epsilon}},
\qquad
\hat{\pi}_i=\frac{p_i}{N},
\label{eq:teacher_distribution}
\end{equation}
and align them through
\begin{equation}
\mathcal L_{\mathrm{sens}}
=D_{\mathrm{KL}}(\boldsymbol{\pi}\|
\hat{\boldsymbol{\pi}})
=\sum_{i=1}^{N}\pi_i
\log\frac{\pi_i}{\hat{\pi}_i}.
\label{eq:sensitivity_loss}
\end{equation}

The teacher distribution is detached in~(\ref{eq:sensitivity_loss}) to avoid second-order differentiation. It is removed after training; inference uses only the token input, the predictor, and the deterministic power shaping in~(\ref{eq:tspa_power})--(\ref{eq:average_power}). TSPA thus distills downstream HR-reconstruction sensitivity into inference-time token power coefficients under the fixed symbol and average-power budgets. As illustrated in Fig.~\ref{fig:tspa}, TSPA produces a nonuniform mean-one power distribution that emphasizes structurally informative regions while retaining protection for reconstruction-relevant non-edge areas, suggesting that its allocation captures useful edge and texture cues without treating them as the sole importance criterion.

\begin{figure*}[!t]
    \centering
    \includegraphics[width=0.84\textwidth]{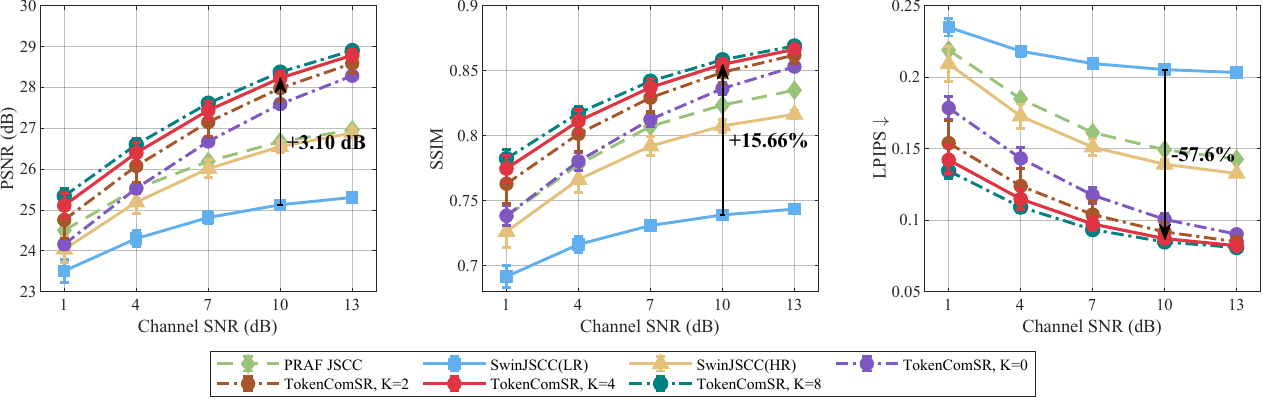}
    \caption{PSNR, SSIM, and LPIPS versus channel SNR at CBR\,=\,1/8 on DIV2K. TokenComSR is evaluated at the trained linear Rician factors $K\in\{0,2,4,8\}$, while the baselines are evaluated at $K=4$. Higher PSNR/SSIM and lower LPIPS indicate better performance.}
    \label{fig:main_results}
\end{figure*}
\subsection{Token Refinement Module}

After wireless transmission, the received tokens may contain residual noise and distortions in their cross-token relationships. Leaving these perturbations entirely to $\mathcal{D}_{\theta}$ would require the decoder to simultaneously invert source compression and suppress channel distortion, while the subsequent nonlinear decoding and SR operations may further amplify the remaining errors. We therefore insert TRM before image decoding to estimate an SNR-conditioned residual directly in token space.

Structurally, TRM operates entirely on the received noisy latent tokens. TRM first projects them into a refinement space through a linear layer,
\begin{equation}
\mathbf{U}_0=\mathbf{W}_{\mathrm{in}}\hat{\mathbf{Z}}
+\mathbf{b}_{\mathrm{in}},
\label{eq:trm_proj_in}
\end{equation}
where $\mathbf{U}_0\in\mathbb{R}^{N\times d_t}$ denotes the hidden representation. The projected tokens are then processed by $L$ stacked Transformer refinement blocks,
\begin{equation}
\mathbf{U}_{\ell}=\mathcal{B}_{\ell}
\bigl(\mathbf{U}_{\ell-1},\mathbf{c}_{\rho}\bigr),
\qquad
\ell=1,2,\ldots,L,
\label{eq:trm_blocks}
\end{equation}
where $\mathbf{c}_{\rho}$ is a conditioning embedding generated from the nominal SNR. We use $L=2$ as a fixed implementation setting in all experiments.

Each refinement block adopts an adaLN-Zero design, in which $\mathbf{c}_{\rho}$ modulates the self-attention and feed-forward paths through shift, scale, and zero-initialized residual gates. The zero-initialized gates make the residual branch close to zero at training onset, so that the initial TRM mapping approximates an identity transformation. After the refinement blocks, the hidden representation is normalized, projected back to the original token space, and added to the received tokens through a residual connection,
\begin{equation}
\bar{\mathbf{Z}}
=
\hat{\mathbf{Z}}
+
\mathbf{W}_{\mathrm{out}}
\operatorname{Norm}
\bigl(\mathbf{U}_{L};\mathbf{c}_{\rho}\bigr)
+
\mathbf{b}_{\mathrm{out}}.
\label{eq:trm_output}
\end{equation}

TRM is optimized under the final SR task objective rather than through direct supervision from clean latent tokens. It therefore performs explicit pre-decoding token-domain residual refinement, while global self-attention exploits cross-token context to restore relationships disturbed by the channel.

\subsection{SR Reconstruction Module in TokenComSR}

Once the refined tokens are decoded into an intermediate LR reconstruction, the bandwidth-compensation task is completed by the SR reconstruction module $\mathcal{R}_{\psi}$. The module comprises a shallow feature extraction layer, multiple stacked residual Swin-T blocks (RSTBs), and a sub-pixel convolution-based upsampling layer. The RSTBs allow the structural information preserved in the decoded feature maps to be further exploited for synthesizing the missing high-frequency details.

The features extracted by the shallow layer are passed through $N_B$ stacked RSTBs. Each RSTB contains multiple Swin-T layers followed by a convolutional layer with residual connections, combining local spatial features with global Transformer interactions. The deep features are then mapped to the target HR output $\hat{\mathbf{I}}_{\mathrm{HR}}$ through a pixel-shuffle upsampling module, which expands the spatial dimensions by the upscaling factor $r$ with minimal computational overhead.
\begin{table*}[!t]
\centering

\caption{Generalization performance of four models across standard SR benchmarks at $\times2$ and $\times4$. All methods use identical channel usage at each scale, with CBRs of $1/8$ and $1/32$ for $\times2$ and $\times4$, respectively. The best and second-best results are shown in \bestresult{red} and \secondresult{blue}.}
\label{tab:full_sr_benchmark_x2_x4}
\small
\setlength{\tabcolsep}{2.4pt}
\renewcommand{\arraystretch}{0.83}
\begin{tabularx}{0.9\textwidth}{@{\extracolsep{\fill}}c l Y Y Y Y Y Y@{}}
\toprule
Scale & Method & Metric & Set5\cite{ref_set5} & Set14\cite{ref_set14} & Kodak24 & Urban100\cite{ref_urban100} & BSD100\cite{ref_bsd100} \\
\midrule
\multirow{12}{*}{$\times 2$} & \multirow{3}{*}{TokenComSR} & PSNR & \bestresult{31.276} & \bestresult{28.067} & \bestresult{28.398} & \bestresult{27.101} & \bestresult{27.174} \\
 &  & SSIM & \bestresult{0.8869} & \bestresult{0.8275} & \bestresult{0.8423} & \bestresult{0.8593} & \bestresult{0.8223} \\
 &  & LPIPS & \bestresult{0.0588} & \bestresult{0.0952} & \bestresult{0.0874} & \bestresult{0.0723} & \bestresult{0.1192} \\
\cmidrule(lr){2-8}
 & \multirow{3}{*}{SwinJSCC-LR} & PSNR & 28.631 & 25.260 & 25.424 & 23.162 & 24.457 \\
 &  & SSIM & 0.8230 & 0.7272 & 0.7253 & 0.7289 & 0.6971 \\
 &  & LPIPS & 0.1147 & 0.1889 & 0.2240 & 0.1907 & 0.2701 \\
\cmidrule(lr){2-8}
 & \multirow{3}{*}{SwinJSCC-HR} & PSNR & \secondresult{29.882} & \secondresult{26.524} & \secondresult{26.718} & \secondresult{24.514} & \secondresult{25.913} \\
 &  & SSIM & \secondresult{0.8539} & \secondresult{0.7879} & \secondresult{0.7916} & \secondresult{0.7903} & \secondresult{0.7792} \\
 &  & LPIPS & \secondresult{0.0639} & \secondresult{0.1184} & \secondresult{0.1489} & \secondresult{0.1417} & \secondresult{0.1780} \\
\cmidrule(lr){2-8}
 & \multirow{3}{*}{PRAF} & PSNR & 27.294 & 24.009 & 24.368 & 22.060 & 23.752 \\
 &  & SSIM & 0.7659 & 0.6741 & 0.6860 & 0.6599 & 0.6552 \\
 &  & LPIPS & 0.2217 & 0.3315 & 0.4078 & 0.3786 & 0.4448 \\
\midrule
\multirow{12}{*}{$\times 4$} & \multirow{3}{*}{TokenComSR} & PSNR & \bestresult{26.954} & \bestresult{23.593} & \bestresult{24.386} & \bestresult{22.694} & \secondresult{23.890} \\
 &  & SSIM & \bestresult{0.7800} & \bestresult{0.6502} & \bestresult{0.6663} & \bestresult{0.6919} & \secondresult{0.6309} \\
 &  & LPIPS & \secondresult{0.1787} & \secondresult{0.2529} & \bestresult{0.2587} & \bestresult{0.1969} & \bestresult{0.2961} \\
\cmidrule(lr){2-8}
 & \multirow{3}{*}{SwinJSCC-LR} & PSNR & 24.171 & 21.582 & 22.792 & 20.407 & 22.477 \\
 &  & SSIM & 0.6842 & 0.5551 & 0.5822 & 0.5681 & 0.5480 \\
 &  & LPIPS & 0.3292 & 0.4334 & 0.4884 & 0.3982 & 0.5261 \\
\cmidrule(lr){2-8}
 & \multirow{3}{*}{SwinJSCC-HR} & PSNR & \secondresult{26.646} & \secondresult{23.404} & \secondresult{24.344} & \secondresult{22.208} & \bestresult{23.933} \\
 &  & SSIM & \secondresult{0.7594} & \secondresult{0.6450} & \secondresult{0.6638} & \secondresult{0.6729} & \bestresult{0.6379} \\
 &  & LPIPS & \bestresult{0.1540} & \bestresult{0.2492} & \secondresult{0.2796} & \secondresult{0.2585} & \secondresult{0.3110} \\
\cmidrule(lr){2-8}
 & \multirow{3}{*}{PRAF$^{*}$} & PSNR & -- & -- & -- & -- & -- \\
 &  & SSIM & -- & -- & -- & -- & -- \\
 &  & LPIPS & -- & -- & -- & -- & -- \\
\bottomrule
\end{tabularx}
\par\vspace{1pt}
{\footnotesize\raggedright
\noindent $^{*}$PRAF has no publicly available official implementation. We reproduce only the $\times2$ configuration reported in the original paper and do not introduce an unverified $\times4$ extension; hence, the dashes denote not applicable.\par}
\end{table*}

\begin{figure}
    \centering
    \includegraphics[width=0.92\linewidth]{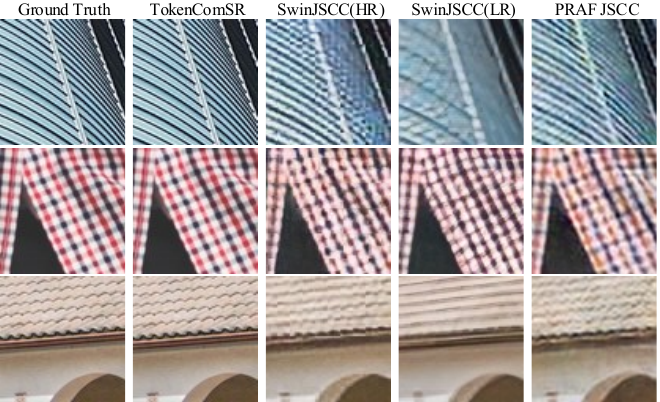}
    \caption{Visual comparison of TokenComSR with baselines at SNR\,=\,10\,dB.}
    \label{fig:visual_results}
\end{figure}

\subsection{Progressive Training}
\label{sec:training}

Let $\mathcal L_{\mathrm{LR}}$ and $\mathcal L_{\mathrm{HR}}$ denote the pixelwise $L_1$ losses at the LR and HR outputs, respectively. $\mathcal L_{\mathrm{per}}$ denotes the HR-domain VGG-feature loss, and $\mathcal L_{\mathrm{sens}}$ is defined in~(\ref{eq:sensitivity_loss}). At phase $s$, the active parameters $\boldsymbol{\Theta}_s$ are optimized by
\begin{equation}
\begin{aligned}
\min_{\boldsymbol{\Theta}_s}\quad
&\mathbb E\!\left[\mathcal L^{(s)}\right],\\
\mathcal L^{(s)}={}&
\lambda_{\mathrm{LR}}^{(s)}\mathcal L_{\mathrm{LR}}
+\lambda_{\mathrm{HR}}^{(s)}\mathcal L_{\mathrm{HR}}+\lambda_{\mathrm{per}}^{(s)}\mathcal L_{\mathrm{per}}
+\lambda_{\mathrm{sens}}\mathcal L_{\mathrm{sens}} .
\end{aligned}
\label{eq:unified_training_objective}
\end{equation}
The expectation is taken over the training images and channel realizations. No additional power penalty is required because~(\ref{eq:tspa_power}) directly enforces the mean-one constraint.

Let $\boldsymbol{\lambda}^{(s)}
=(\lambda_{\mathrm{LR}}^{(s)},\lambda_{\mathrm{HR}}^{(s)},
\lambda_{\mathrm{per}}^{(s)})$.
In Phase~1, $\boldsymbol{\lambda}^{(1)}=(1,0,0)$; the encoder, decoder, and TSPA predictor are trained, while TRM is bypassed and SR is frozen. In Phase~2, $\boldsymbol{\lambda}^{(2)}=(0.1,1,0.05w_{\mathrm p})$; the decoder, TRM, and predictor are trained with the encoder and SR frozen. Here, $w_{\mathrm p}$ is zero for the first 50 Phase-2 epochs and linearly increases to one over the next 50 epochs. In Phase~3, $\boldsymbol{\lambda}^{(3)}=(0.1,1,0.05)$ and SR is additionally unfrozen. Phase~4 restores the best Phase-3 model and fine-tunes all modules with $\boldsymbol{\lambda}^{(4)}=(0.1,1,0.05)$ and reduced learning rates. The task-sensitivity weight $\lambda_{\mathrm{sens}}$ is set to $0.05$ during the first half of Phase~2 and then linearly annealed to zero over the remaining Phase-2 epochs; it remains zero in the subsequent phases.

\section{Experimental Results}
\label{sec:experiment}
The performance of the proposed TokenComSR framework is evaluated through comprehensive numerical simulations. TokenComSR is compared against state-of-the-art JSCC schemes with cascaded SR and recent JSCC-SR system.

\subsection{Experimental Settings}
\label{sec:experimental_settings}

TokenComSR is trained and evaluated on the DIV2K dataset \cite{ref_div2k}, split into 650\,:\,75\,:\,75 images for training, validation, and testing. Random $64\times64$ LR crops and the corresponding $128\times128$ HR targets are used for training. During training, $\rho$ is uniformly sampled from [1,13] dB, while $K$ is uniformly selected from $\{0,2,4,8\}$. Evaluation uses several $\rho$ values and the same four Rician factors. For each CBR setting, we independently train a dedicated model with the corresponding channel dimension. The inference latency is measured on NVIDIA RTX 5090 under the same hardware environment.

The baselines include PRAF-JSCC \cite{jscc_sr} and two SwinJSCC \cite{swinjscc} variants: SwinJSCC-LR, which appends a pretrained Swin2SR module after the decoder for HR reconstruction, and SwinJSCC-HR, which applies Swin2SR at the transmitter and communicates an HR representation. As the advantages of Deep JSCC over conventional separate source-channel coding under bandwidth-limited fading channels have been extensively established in prior studies, we omit this well-established comparison and focus on more competitive learning-based JSCC baselines. All schemes are evaluated under identical conditions for each operating point, across multiple CBRs and Rician K-factors. Communication quality is measured against the ground-truth HR images using PSNR \cite{ref_psnr}, SSIM \cite{ref_ssim}, and LPIPS \cite{ref_lpips}; higher PSNR/SSIM and lower LPIPS indicate better performance.
\begin{figure*}[!t]
    \centering
    \includegraphics[width=0.84\textwidth]{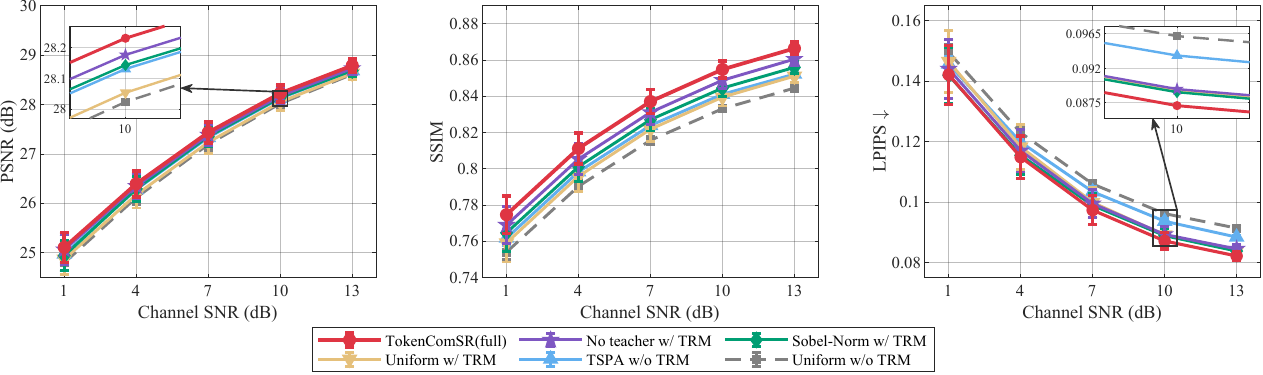}
    \caption{Ablation study of different power-allocation strategies and TRM configurations in terms of PSNR, SSIM, and LPIPS at CBR\,=\,1/8. Error bars denote the sample standard deviation over five paired channel trials.}
    \label{fig:ablation}
\end{figure*}

\subsection{Performance Evaluation}

Fig.~\ref{fig:main_results} compares PSNR, SSIM, and LPIPS versus channel SNR at CBR\,=\,1/8. The baselines are evaluated at linear $K=4$, while TokenComSR is additionally tested at $K\in\{0,2,4,8\}$ in accordance with its multi-$K$ training protocol. All methods use the same channel-use budget and follow the same $\times2$ and per-image Rician block-fading protocol at each operating point. The performance of all schemes improves smoothly with SNR, without an abrupt quality collapse. For TokenComSR, increasing $K$ from the Rayleigh limit at $K=0$ to $K=8$ consistently improves PSNR and SSIM and reduces LPIPS, reflecting the reduced fading severity as the LoS component becomes stronger.

Among the compared methods, TokenComSR achieves the highest PSNR and SSIM and the lowest LPIPS across the evaluated SNR range. At SNR\,=\,10\,dB, it obtains 28.2305\,dB PSNR, 0.85489 SSIM, and 0.08715 LPIPS. Relative to the SwinJSCC-LR baseline, it improves PSNR by 3.10\,dB and SSIM by 15.66\%, while reducing LPIPS by 57.6\%. It also exceeds the SwinJSCC-HR and PRAF JSCC baselines by 1.68\,dB and 1.56\,dB in PSNR, respectively. The visual comparison at SNR\,=\,10\,dB in Fig.~\ref{fig:visual_results} further confirms this perceptual advantage, showing that TokenComSR better preserves task-relevant structures and important visual details.

At an actual SNR of 10~dB, conditioning offsets within $\pm3$~dB cause a total PSNR variation of only 0.0270~dB and a maximum degradation of 0.0143~dB relative to matched conditioning, indicating low sensitivity to moderate nominal-SNR errors. To evaluate different channel-use budgets, all schemes are independently trained and tested at CBRs of $1/16$, $1/8$, and $1/4$ under $K=4$ and SNR\,=\,10\,dB. TokenComSR achieves PSNR values of 26.90, 28.23 and 29.34~dB, respectively, outperforming the best baseline by 1.30--1.68~dB across the three CBRs. Consistent advantages are also observed in SSIM and LPIPS, confirming that the performance gain persists across different channel-use budgets.

\begin{table}[!t]
\caption{Computational Complexity and Inference Latency of TokenComSR and the SwinJSCC Baselines}
\label{tab:complexity}
\centering
\renewcommand{\arraystretch}{0.94}
\resizebox{\columnwidth}{!}{%
\begin{tabular}{ccccc}
\hline
\textbf{Method} & \textbf{Side} & \textbf{Params. (M)} & \textbf{GMACs} & \textbf{Latency (ms)} \\
\hline
             & Tx    & 14.315 & 1.108  & 4.562 \\
SwinJSCC-LR  & Rx    & 26.403 & 53.126 & 24.397 \\
             & Total & 40.718 & 54.234 & 28.896 \\
\hline
             & Tx    & 26.226 & 56.433 & 24.168 \\
SwinJSCC-HR  & Rx    & 14.131 & 4.416  & 5.063 \\
             & Total & 40.357 & 60.849 & 29.422 \\
\hline
             & Tx    & 14.052 & 1.704  & 8.859 \\
TokenComSR   & Rx    & 23.666 & 53.752 & 29.442 \\
             & Total & 37.718 & 55.455 & 38.591 \\
\hline
\end{tabular}%
}
\end{table}
\subsection{Cross-Dataset Generalization at Different SR Scales}

For each SR scale, we evaluate cross-dataset generalization on Set5~\cite{ref_set5}, Set14~\cite{ref_set14}, Kodak24, Urban100~\cite{ref_urban100}, and BSD100~\cite{ref_bsd100} without dataset-specific retraining. As reported in Table~\ref{tab:full_sr_benchmark_x2_x4}, all methods use identical channel usage, corresponding to HR-referenced CBRs of $1/8$ and $1/32$ for $\times2$ and $\times4$, respectively. The results are averaged over five paired channel realizations under per-image Rician block fading at SNR\,=\,10~dB and $K=4$. All metrics are computed on RGB images without border shaving using deterministic $64\times64$ LR crops. 

At $\times2$, TokenComSR ranks first in all three metrics across the five benchmarks. Its consistent advantage over SwinJSCC-HR, the strongest competing baseline, covers natural images, texture-rich content, and urban scenes, indicating that the learned task-sensitive protection and token refinement generalize beyond the DIV2K test distribution rather than favoring a particular content type. The $\times4$ setting is more challenging because of its more ill-posed reconstruction and fourfold lower HR-referenced channel-use budget. TokenComSR achieves the best PSNR and SSIM on four datasets and the best LPIPS on three, while outperforming SwinJSCC-HR in all three dataset-averaged metrics. Overall, TokenComSR provides consistently strong results across all three evaluation dimensions under the more demanding $\times4$ setting.

\subsection{Ablation Study and Complexity Analysis}
\label{sec:ablation}
Six independently trained variants are evaluated in Fig.~\ref{fig:ablation}. The four configurations formed by TSPA or uniform power allocation with and without TRM constitute a $2\times2$ module-level ablation, isolating the effects of transmitter-side power allocation and receiver-side token refinement. We further include Sobel-based allocation and a teacher-free learned allocator, both with TRM enabled, to examine whether the gain of TSPA arises from the allocation criterion itself or merely from introducing a learned power predictor. The complete TokenComSR achieves the best overall reconstruction fidelity, confirming the complementary roles of TSPA and TRM.

Both TSPA and TRM provide consistent, albeit modest, improvements in PSNR and SSIM over the uniform/no-TRM baseline. Their combination yields a clearer perceptual benefit, reducing the five-SNR average LPIPS by 0.0085, of which TRM alone contributes a reduction of 0.0057. Under the matched TRM-enabled comparison, TSPA consistently outperforms the teacher-free allocator in all three metrics. It also achieves lower LPIPS than Sobel at four SNRs and is nearly tied at 1~dB. These results confirm the complementary roles of task-sensitive power allocation and receiver-side token refinement, particularly in improving perceptual reconstruction quality.

Table~\ref{tab:complexity} compares model parameters, multiply-accumulate operations (MACs), and inference latency. TokenComSR uses fewer parameters than both SwinJSCC baselines, with arithmetic complexity comparable to SwinJSCC-LR and lower than SwinJSCC-HR, indicating that its gains do not arise merely from increased model scale. Its approximately one-third higher end-to-end latency represents a reasonable trade-off for the achieved reconstruction improvements. Most computation remains at the receiver, consistent with the intended compute-asymmetric edge-to-server architecture.

\section{Conclusion}

This paper presented TokenComSR, an end-to-end TokenCom framework for wireless transmission of LR observations and receiver-side HR reconstruction. TSPA distills downstream HR-reconstruction sensitivity into token-level power coefficients under fixed channel usage and average-power constraints, while TRM performs SNR-conditioned token-domain residual refinement before nonlinear decoding and SR. A four-phase progressive training strategy coordinates the coupled transceiver, token refinement, and SR reconstruction modules. Experiments under per-image Rician block fading demonstrate improvements in reconstruction fidelity and perceptual quality, together with robust performance across the evaluated SNRs, Rician factors, channel-use budgets, datasets, and SR scales. Future work will consider time-selective fading, channel-estimation errors, and variable-resolution inference.


\bibliographystyle{IEEEtran}
\bibliography{IEEEabrv,reference,reference_revision}

\end{document}